\documentclass[12pt]{article}
\usepackage{latexsym}
\title{The Volterra lattice as a gradient flow}
\author{A. V. Pensko\"\i\thanks{Faculty of Mechanics and Mathematics, 
Department of Higher Geometry and Topology, Moscow State University, 
Vorob'ievy Gory,
Moscow, 119899, Russia. {\tt e-mail: penskoi@mech.math.msu.su}}}
\date{}
\begin{document}
\maketitle
\abstract{The Volterra lattice is considered.
New gradient interpretation for this dynamical system is proposed.
This interpretation seems to be more natural than existing ones.}

\smallskip

\noindent\textbf{AMS MSC 58F19, 58F07}

\bigskip

In the present paper we give a gradient interpretation for the Volterra
lattice treated in details in~\cite{FT}.
The phase space consists of variables
$u_n>0,\, 1\le n\le N.$
The equations of motion are
$$
\frac{du_n}{dt}=u_n(u_{n+1}-u_{n-1}),
$$
where $u_0=u_{N+1}=0.$

The interest to gradient interpretations goes back to the paper by
J.~Moser~\cite{M},
where such an interpretation was given implicitly
for the Toda lattice.
Twenty years  later A.~Bloch, R.~Brockett and T.~Ratiu~\cite{BBR}
suggested another gradient interpretation for the Toda lattice,
which was different from Moser's one.
Our interpretation for the Volterra lattice as a gradient flow
is based on the same idea of the double bracket representation
(see~\cite{BBR}).

It should be noted that as well as for the Toda lattice
our interpretation is not unique: the well-known isomorphism between these two
problems (see e.g.~\cite{D}) and the results of~\cite{BBR}
lead to one more gradient interpretation for the
Volterra lattice.
Our representation seems to be more natural.
As a remark we note that the gradient interpretation
for the Toda lattice was used in the paper
by C.~Tomei~\cite{T} and D.~Fried~\cite{F} for study
of non-trivial geometry of isospectral sets.

Let $c_i=\sqrt{u_i}$.
Let us consider a manifold $M$ of all matrices of the form
$$
L=\left(%
\begin{array}{ccccc}
0 & c_1 & 0 & \dots & 0 \\
c_1 & 0 & c_2&  & \vdots \\
0 & c_2 & 0 & \ddots  & \vdots \\
\vdots & & \ddots  & 0 & c_N \\
0 & \dots & \dots    & c_N  & 0 \\
\end{array}\right)
$$
with a given spectrum.
The manifold $M$ is called the isospectral set.
Let $V$ be the euclidean space of $(N+1)\times(N+1)$-matrices
with respect to the scalar product
$\langle X,Y\rangle=\mbox{tr}XY^{T}.$
The manifold $M$ is naturally embedded in $V$.
Let $L_0$ be any point of $M$.
Consider the adjoint $GL_{N+1}$-action in $V$; then
$GL_{N+1}L_0$ is a submanifold of $V$ and $M$ is a submanifold
of $GL_{N+1}L_0$.
Let $V_{L_0}=\{T\in V|[L_0,T]=0\},$
let $V_{L_0}^{\perp}$ be its orthogonal
complement
with respect to $\langle , \rangle$; then
$V_{L_0}^{\perp}\cong T_{L_0}GL_{N+1}L_0.$
By $T^{\perp}$ denote the orthogonal projection
of $T\in V$ on $V_{L_0}^{\perp}.$
Define the scalar product of two vectors
$[L_0,A], [L_0,B]\in T_{L_0}GL_{N+1}L_0$ as
$([L_0,A],[L_0,B])=\langle A^{\perp}, B^{\perp}\rangle.$
In this way we obtain
a Riemannian metric on $GL_{N+1}L_0$. Its restriction gives a
Riemannian metric on $M$.

Let
$V\ni K=\frac{1}{4}\mbox{diag}(1,2,3,4,\dots)$ and
$f(L)=\langle K,L^2\rangle=\mbox{tr}KL^2.$

{\noindent\sc Theorem.}
The gradient flow of
$f|_M$ relative to the Riemannian metric on $M$ coincides with the
restriction of Volterra flow to a joint level of its integrals.

{\noindent\sc Proof.}
It is known (see e.g.~\cite{D}) that
the Volterra lattice is equivalent to the
Lax equation
$\dot{L}=[L,A],$
where $A$ is the matrix
$$
\left(%
\begin{array}{ccccccc}
0 & 0 & \frac{1}{2}c_1c_2 & 0 & \dots &  & 0 \\
0 & 0 & 0 & \frac{1}{2}c_2c_3 & & & \vdots \\
-\frac{1}{2}c_1c_2 & 0 & 0 & & \ddots & & \\
0 & -\frac{1}{2}c_2c_3 & & & & & 0 \\
\vdots & & \ddots & & & & \frac{1}{2}c_{N-1}c_N \\
  & & & & & 0 & 0\\
0 & \dots & & 0 & -\frac{1}{2}c_{N-1}c_N & 0 & 0 \\
\end{array}\right).
$$
The integrals of motion are $\mbox{tr}L^k$; therefore
$M$ is diffeomorfic to the joint level of integrals.
It is sufficient to prove that
the restriction on $M$ of the gradient flow of
$f|_{GL_{N+1}L}$ relative to the Riemannian metric
$(,)$ coincides with the Volterra flow.

Let $[L,T]\in T_LGL_{N+1}L$ be an arbitrary vector.
Let us consider $X$ such that $[L,X]=\mbox{grad}(f|_{GL_{N+1}L})$;
then we have the equation
$df([L,T])=([L,X],[L,T]).$
We see that
$df([L,T])=\langle K,L[L,T]+[L,T]L\rangle=\langle K,[L^2,T]\rangle=%
\langle[L^2,K],T\rangle.$
Further, $[L^2,K]^{\perp}=[L^2,K],$
therefore $df([L,T])=\langle[L^2,K]^{\perp},T\rangle.$
At the same time
$([L,X],[L,T])=\langle X^\perp, T^\perp\rangle$;
then $X^{\perp}=[L^2,K]$ and $\mbox{grad}(f|_{GL_{N+1}L})=[L,[L^2,K]].$

On the other hand it is easy to prove that $A=[L^2,K],$
thus the Volterra lattice is equivalent to the equation
$\dot{L}=[L,[L^2,K]].$ This completes the proof.
$\Box$

\end{document}